\documentclass[%
 reprint,
floatfix,
 amsmath,amssymb,
 aps,
]{revtex4-1}

\usepackage{graphicx}
\usepackage{dcolumn}
\usepackage{bm}
\usepackage{comment}
\usepackage{mathtools}

\begin{document}

\preprint{APS/123-QED}

\title{The Librator: A new dynamical regime for nonlinear microelectromechanical devices}
\author{Samer Houri}
\email{Samer.Houri.dg@hco.ntt.co.jp}
  \affiliation{NTT Basic Research Laboratories, NTT Corporation,
3-1 Morinosato-Wakamiya, Atsugi-shi, Kanagawa 243-0198, Japan.}

\author{Motoki Asano}
\author{Hajime Okamoto}
\author{Hiroshi Yamaguchi}
  \affiliation{NTT Basic Research Laboratories, NTT Corporation,
3-1 Morinosato-Wakamiya, Atsugi-shi, Kanagawa 243-0198, Japan.}
\date{\today}
\renewcommand{\vec}[1]{\mathbf{#1}}
\begin{abstract}
We present a novel mode of operation for Duffing-type nonlinear microelectromechanical (MEMS) devices whereby a self-sustained multi-frequency 
output is generated. This new librator regime creates a limit cycle around a dynamical fixed point, i.e. around fixed points within the rotating frame, whereas a traditional oscillator generates a limit cycle around a static fixed point. The librator limit cycles thus created do not change the global topology of the rotating frame phase space, but are constrained by it. Due to the Duffing nonlinearity different types of limit cycles could be generated within the same phase space, with each type possessing distinct dynamical features. Transitioning between these limit cycles requires crossing homoclinic bifurcations, which is done without generating chaos as the phase space dynamics are two dimensional. This work opens the possibility to the creation of a librator network in analogy with oscillator network, however this can be done in a single MEMS device.
\end{abstract}

\maketitle
\section{\label{sec:level1}Introduction}
\indent\indent\ Self-oscillating systems (or simply oscillators), defined as systems that produce a periodic output without being periodically driven \cite{jenkins2013self}, are omnipresent in the physical \cite{pippard2007physics}, biological \cite{glass2001synchronization,glass2020clocks}, and engineering fields \cite{van1927vii}. Indeed, oscillators are at the center of modern electronic instruments as they provide frequency and timing references \cite{vittoz2010low}. Microelectromechanical and nanoelectromechanical (M/NEMS) devices in particular represent an interesting medium for the realization of self-oscillating systems as they provide high quality factors, low-power operation, and on-chip integration capabilities \cite{van2011review}.\\
\indent\indent\ However, beyond their time keeping role, oscillators are also crucial for the study of complex phenomena that arise due to their coupling such as synchronization \cite{pikovsky2003synchronization}, chimeras \cite{abrams2004chimera}, and phase patterns \cite{kuramoto2003chemical}. Furthermore, large networks of coupled oscillators form the building blocks for new computational techniques such as neuromorphic computing \cite{torrejon2017neuromorphic} and reservoir computation \cite{tanaka2019recent}.\\
\indent\indent In dynamical terms self-sustained oscillators are described as limit cycles, which are closed attracting orbits in the corresponding phase space \cite{strogatz2018nonlinear}. 
The phase space itself is a two dimensional representation of the fixed points and the vector fields that govern the dynamics of a system. Therefore, as the number of fixed points in the phase space increases, and the associated vector fields take on more elaborate forms, the dynamics of the system will also become richer and more varied, with the potential for hitherto new dynamics \cite{takens2001forced}.\\
\indent\indent\ In their most common physical implementation, the van der Pol oscillator \cite{van1927vii}, limit cycles possess an amplitude independent frequency and a purely circular trajectory orbiting an unstable fixed point in the phase space. Although more elaborate orbits maybe obtained from the van der Pol oscillator \cite{van1926lxxxviii,le1960two}, its dynamics remain limited by the topology of the phase space in which it is contained. Here, topology refers to the number and nature of the fixed points present within the phase space.\\
\indent\indent\ Thus, a means to increase the gamut of the dynamical response of an oscillator would be to change its phase space topology, i.e. increase the number of fixed points. However, controlling the phase space topology in physical devices is far from trivial and usually requires a complete system re-design, if at all possible. For instance, if one wishes to increase the number of fixed points of the phase space of a MEMS device from a single fixed point, as is the case for a van der Pol oscillator, to 3 fixed points, would require the creation of a static double well potential which would imply the use of specially designed devices and materials \cite{demartini2007chaos,park2008energy,de2006complex,charlot2008bistable}. Such difficulties could explain the dearth of experimental investigation of double-well self-oscillators with only modelling efforts undertaken \cite{alhussein2020potential,datta2017bifurcations}.\\
\indent\indent\ Luckily, nonlinear M/NEMS resonators offer a means to overcome such experimental constraints through the use of their dynamical properties, since M/NEMS resonators 
can have their dynamics treated using the rotating frame approximation (RFA). The RFA permits the separation of scale and the averaging out of fast oscillations, leaving a small envelope around a resonance mode in which the dynamics take place. Those rotating frame dynamics can in their turn be embedded in a 
rotating frame phase space, which would sustain multiple fixed points and a more interesting topology for a device that would otherwise have a single fixed point in its laboratory frame phase space. For instance, a weakly nonlinear M/NEMS resonator, that exhibits a single fixed point at rest, demonstrates three fixed points when driven directly or parametrically \cite{dykman1998fluctuational,tadokoro2020noise,dolleman2019high,mahboob2008bit}.\\
\indent\indent\ Since the RFA phase space can be manipulated by simply changing the experimental conditions (usually by modifying a driving force), fixed points can therefore be created and manipulated without requiring device or setup redesign. This experimental flexibility explains the recent interest in the RFA dynamics of M/NEMS nonlinear devices, where rotating frame dynamics were used to demonstrate noise squeezing \cite{huber2020spectral}, chaos \cite{houri2020generic}, as well as solitons \cite{yamaguchi2021generation} and pseudo-angular momenta systems \cite{asano2019optically}.\\
\indent\indent\ In this work we expand on the current inventory of experimental nonlinear dynamical systems, particularly as they apply to microelectromechanical systems (MEMS) ones, by introducing a new regime in which a limit cycle is created around the fixed points within the rotating frame phase space. The limit cycles thus observed demonstrate a richer dynamical behaviour when compared to the limit cycles in case of conventional oscillators. We label these rotating frame limit cycles ``librators", and model and experimentally explore their behaviour.\\
\indent\indent Libration is a term used in the context of celestial mechanics to indicate periodic motion around a dynamical fixed point, for example when a spacecraft orbits one of the Lagrangian points in the earth-moon or sun-earth rotating frame \cite{cassini,farquhar1970control,howell1997application}. And although the term libration is sometimes used 
to designate a different dynamical aspect \cite{strogatz2018nonlinear,alexeeva2000impurity,lenz2007classical,nakajima1994orientational,rost1992pendular,friedrich1995alignment}
, we employ the term in this work to only indicate periodic orbits around dynamical fixed points.\\
\indent\indent\ Furthermore, we extend the concept of libration orbits and use the term librator to indicate a limit cycle created around a dynamical fixed point (i.e. in the rotating frame) of a microelectromechanical system, in analogy with the use of the terms oscillation and oscillator. The distinction between librator and oscillator being that whereas the former's limit cycle is created around a dynamical fixed point, the latter's is created around a static fixed point, usually the rest position. Therefore, the output of a librator as seen in the laboratory frame is (nearly) quasi-periodic, whereas that of an oscillator is periodic.\\
\indent\indent\ As will be shown below, the librator affords access to highly unusual and interesting dynamics; for instance, the creation, in a controllable manner, of a rotating frame phase space that supports 
distinct types of structurally stable 
limit cycles which are separated by homoclinic bifurcations
. More importantly, since the dynamics of the librator are fully contained in a two-dimensional phase space, the crossing of the homoclinic does not result in the onset of chaos, thus enabling new exotic and chaos-less dynamics. To the best knowledge of the authors, this is the first experimental demonstration of a controllable homoclinic bifurcation in a microelectromechanical system, despite previous interesting indications of transient critical slowing down in ring-down measurements \cite{bagheri2011dynamic}.\\
\begin{figure}
	\graphicspath{{Figures/}}
	\includegraphics[width=85mm]{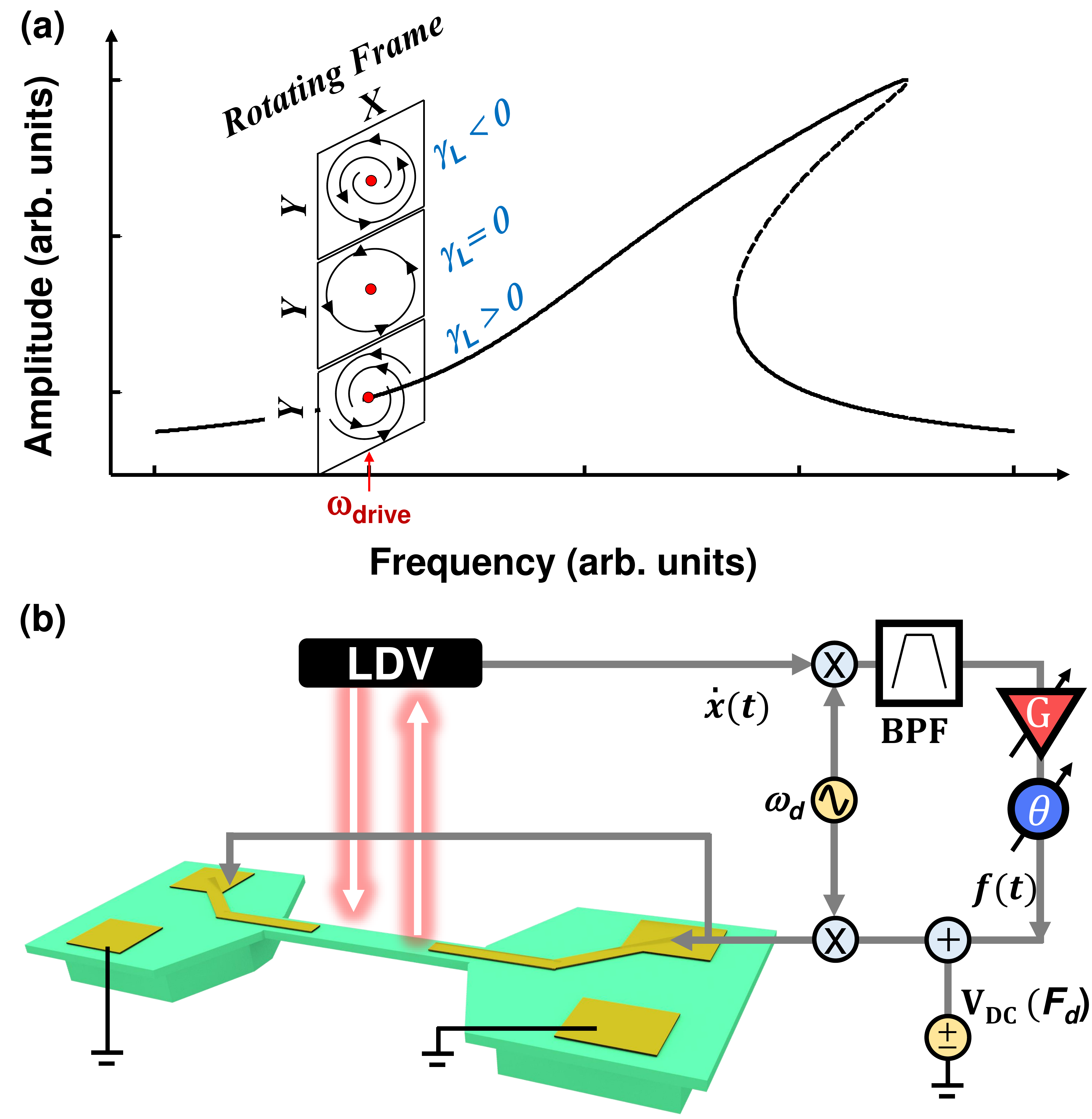}
	\caption{(a) Schematic representation of a librator. A nonlinear resonator driven outside the bistable regime has a single steady-state solution, i.e. fixed point, in the rotating frame phase-space (red dots). A perturbation around the steady-state results in a transient response in which the system slowly spirals back towards the fixed point (bottom inset). If the libration damping ($\gamma_L$) is zero, i.e. $\gamma_L=0$, then the perturbation will persist as a periodic motion in the rotating frame phase space orbiting the fixed point (middle inset). if the libration damping is negative, i.e. $\gamma_L<0$, the fixed point becomes unstable and leads to a limit cycle being created in the rotating frame phase space (top inset). (b) Feedback loop for the creation of a librator. The structure's motion is measured using a laser Doppler vibrometer (LDV), its output ($\dot{x}(t))$ is then down-converted, band-pass filtered (BPF), amplified (G) and phase shifted ($\theta$), before adding a DC voltage component that corresponds to the steady state forcing ($F_d$), and then upconverted to the drive frequency ($\omega_d$) and injected to drive the MEMS device.}
\end{figure}
\section{\label{sec:level1}Theory and Modelling}
\indent\indent\ To properly introduce the concept of librator, we start by considering a driven weakly nonlinear Duffing-type MEMS resonator, and to simplify matters further we consider the resonator to be driven outside the hysteretic region, i.e. outside the bistable region, as shown in Fig.~1(a). Such system will have a steady state response, i.e. a vibration amplitude, dictated by its parameters and those of the external forcing. If the system is perturbed from its steady state, it will undergo a transient oscillation on top of the drive oscillation as it returns to its original amplitude \cite{antoni2013nonlinear,zaitsev2012nonlinear}. Seen in the rotating-frame phase space, this transient corresponds to the system tracing a spiral as it approaches the steady-state fixed point as shown in the inset of Fig.~1(a). The situation becomes more interesting if the libration motion is undamped, in such a case the small perturbation will persist and will continuously orbit the fixed point in the rotating frame, resulting in an undamped libration oscillation, also shown in Fig. ~1(a). This concept maybe further extended by the deliberate creation of a limit cycle around the rotating-frame fixed point, in such a situation the originally attracting fixed point becomes a repeller and a stable limit cycle is created around the, now unstable, fixed point. Thus libration oscillations are now self-sustained and the system is a librator, also shown in Fig.~1(a).\\
\indent\indent\ Driving a nonlinear resonator with sinusoidal forcing results in a steady-state solution that is represented by a fixed point within the rotating frame, or two stable fixed points and one saddle point in case the system is driven into the bistable regime. Since the dynamics we seek, i.e. the limit cycle, is equally supposed to take place within the rotating frame, it is clear that simply forcing the device with a sinusoidal drive is not enough, hence additional terms are needed to create the interesting dynamics. Therefore, the standard equation of a driven nonlinear resonator \cite{cleland2013foundations} is modified to include an additional term in the rotating frame, indicated by \emph{f(t)}, the equation now reads\\
\begin{multline}
{\ddot{x} + {(\gamma + \beta x^2)}\dot{x} + \omega_{0}^2x + \alpha x^3 = }{{(F_d + f(t))}\cos(\omega_dt)}\\
\label{eqn:Eq1}
\end{multline}
where $x$ is the displacement, and $\gamma$, $\beta$, $\omega_0$, $\alpha$ are respectively the linear damping, nonlinear damping, natural frequency, and Duffing nonlinearity of the resonator. ${F_d}$ and $\omega_d$ are the amplitude and frequency of the applied external forcing, and \emph{f(t)} is the additional term necessary to create a limit cycle within the rotating frame. Both $F_d$ and \emph{f(t)} will be given in units of Volts throughout this text, however, in order to balance the equation a transduction coefficient $\eta$ is implicitly included in those terms. In addition a detuning parameter $\delta$ is introduced such that $\omega_d = \omega_0\times(1+\delta)$. We also introduce the scaled constants as ${\bar{t}=t\times\omega_0}$, ${\bar{\gamma}=\gamma/\omega_0}$, ${\bar{\alpha}=\alpha/\omega_0^2}$, ${\bar{\beta}=\beta/\omega_0}$, ${\bar{{F_d}}={F_d}/\omega_0^2}$, and ${\overline{f}(t)={f(t)}/\omega_0^2}$. Hereon, all equations are written using this form, however, the bars are dropped for convenience. Note that the terms $\gamma$, $\beta$, $\alpha$, ${F_d}$, $\delta$, and \emph{f(t)} are all perturbation order terms, i.e. ${\sim\mathcal{O}}({\epsilon})$, thus indicating a weakly nonlinear, weakly forced system.\\
\indent\indent\ The usual approach to creating a limit cycle in a MEMS oscillator consists of inserting a resonator in a feedback loop, as the gain of the feedback loop is increased the effective damping of the resonator ($\gamma_{eff}$) is decreased until it becomes negative and a limit cycle is thus created \cite{yurke1995theory,van2013nonlinear,villanueva2013surpassing,chen2016self,ohta2017feedback}. This approach is not suitable for librators 
which impose two conditions. First, since the limit cycle is to be created in the rotating frame, the feedback needs to be applied only in the rotating frame, hence the feedback term (\emph{f(t)}) in Eq.~(1) is multiplied by the driving frequency which acts as a carrier frequency. Second, since the aim is to create a limit cycle around the driven fixed point(s), the response corresponding to the driven term needs to be excluded from the feedback loop so it would not be amplified.\\
\indent\indent\ An implementation of a feedback loop that fulfills these conditions is shown in Fig.~1(b). Wherein, the output of a driven MEMS resonator is measured using a laser Doppler vibrometer (LDV), which measures the velocity. The LDV signal is downconverted using a lock-in amplifier, thus capturing the rotating frame dynamics, and passed through a band-pass filter so as to remove the dc component, which corresponds to the carrier frequency component. The upper cutoff frequency of the filter acts to limit the bandwidth of the feedback loop to within a desired range around the carrier. This output is then amplified, phase-shifted, and used to modulate the carrier frequency thus implementing a gain loop within the rotating frame of the driving force, which at the same time does not change the driven response due to that force (see appendix A for more details on the experimental setup).\\
\indent\indent\ The dynamics of the librator are obtained by analysing the system, including the feedback loop, using the rotating frame approximation (RFA) \cite{cleland2013foundations}, where the displacement $x$ is supposed to take the form ${x(t)=R(t)\cos(\omega_d t+\phi(t))}$, where ${R(t)}$ and ${\phi(t)}$ are slowly varying amplitude and phase envelopes (slow flow variables). We introduce the complex phase space envelop ${A}(t) = Re^{i\phi}$, and its complex conjugate ${A}^*$. In addition, since the motion within the rotating frame consists of a steady-state amplitude component and a superimposed libration component, the slow-flow variables are further decomposed into ${A}(t) = {A_0} + {A_L}(t)$ where ${A_0 = R_0e^{i\phi_0}}$ is the static component
, whereas ${A_L}(t) = R_Le^{i\phi_
L}$ is the time dependent libration component.\\
\indent\indent For low amplitude librations, the libration motion is considered to be centered aournd the steady state component (${A_0}$) which is obtained by solving the standard forced nonlinear resonator equation \cite{cleland2013foundations}. Whereas the dynamics are obtained by developing an expression for the feedback such that $f(t) = f({A(t)},{A(t)}^*)$, inserting it into Eq.~(1) and expanding to give the following governing equations (see Appendix B for the detailed derivation)\\

\begin{widetext}
\begin{eqnarray}
\begin{cases}

{\dot{A}_L=-(i\delta_L + \frac{1}{2}\gamma_L)A_L  +\frac{1}{8}(i3\alpha - \beta)h(A_L)+C_LA^*_L}\\
{\dot{A}^*_L=(i\delta_L - \frac{1}{2}\gamma_L)A^*_L  - \frac{1}{8}(i3\alpha + \beta)h(A_L)^*+C_L^*A_L}
\label{eqn:Eq2}
\end{cases}
\end{eqnarray}
\end{widetext}
where
\begin{eqnarray}
\begin{cases}
{\gamma_L = \gamma + \frac{1}{2}\beta R_0^2 -\frac{1}{4}g\cos\theta}\\
{\delta_L = \delta - \frac{3}{4}\alpha R_0^2 + \frac{1}{8}g\sin\theta}\\
{h(A_L) = 2A_0R_L^2 + A_0^*A_L^2 + R_L^2A_L}\\
{C_L = \frac{1}{8}((i3\alpha-\beta) A_0^2-g e^{i\theta})}
\label{eqn:Eq3}
\end{cases}
\end{eqnarray}
where \emph{g} is the loop gain, $\theta$ is the feedback phase, and $\delta_L$ and $\gamma_L$ are respectively the effective detuning and effective linear damping of the libration motion ${A_L}$. While the quadratic and cubic terms in ${A_L}$ and ${A_L}^*$ are collected in the functions $h({A_L})$ and $h({A_L})^*$, respectively.\\
\indent\indent\ Equation~(2) is a two dimensional autonomous system, which indicates that as long as the RFA is valid and higher order terms can be safely neglected, the system cannot exhibit chaotic dynamics. Note that Eq.~(2) always has $A_L = A_L^* = 0$ as a fixed point, although not necessarily a stable one.\\
\indent\indent\ As the feedback gain term $g$ is increased, the effective libration linewidth $\gamma_{L}$ decreases until reaching zero, for $g\cos\theta = 4\gamma + 2\beta R_0^2$, at which point a libration limit cycle is generated via a Hopf bifurcation. Near the Hopf bifurcation a libration frequency ($\omega_L$) can be obtained by linearizing  Eq.~(2), i.e. dropping the $h({A_L})$  and $h({A_L})^*$ terms, and calculating the eigenvalues of the system, which gives\\
\begin{equation}
\omega_\textrm{L} = \pm \textrm{real}\left[ \sqrt{\delta_L^2 - \vert C_L\vert^2}\right]
\label{eqn:Eq4}
\end{equation}
\indent\indent\ Equation~(4) gives the libration frequency, $\omega_L$, for low amplitude librator limit cycles, i.e. $R_L\approx 0$. If we set $\gamma = \beta = g = 0$, then Eq.~(4) reduces to the libration frequency of a hamiltonian system as given in \cite{ochs2020multiphoton,houri2020generic}.
\section{\label{sec:level1}Experiment and Discussion}
\indent\indent\ Experimental investigation of librator dynamics are conducted using a piezoelectrically actuated GaAs heterostructure MEMS clamped-clamped beam device that is 100 $\mu$m in length, 20 $\mu$m wide, and 600 nm in thickness, see \cite{yamaguchi2017gaas,houri2019limit} for more information on device fabrication. The device is placed in a vacuum chamber with a pressure of $\sim 1$ mPa, excited electrically, and its vibrations measured optically using a LDV. The actuation voltage, which is applied to both electrodes, generates piezoelectric stress in the mechanical resonator that leads to the bending of the resonator due to the built-in layered structure.\\
\indent\indent\ We measure and quantify the main device properties (see appendix A and \cite{houri2019limit,davidovikj2017nonlinear,zaitsev2012nonlinear,polunin2016characterization} for procedures of various parameter fitting) as follows: $\omega_0 = 2\pi\times960$ kHz, quality factor of 1042, and a scaled Duffing nonlinearity of $\alpha=48$. We place the device in a feedback loop that is functionally equivalent to that shown in Fig.~1(b), and measure its response while the drive terms ($F_d$ and $\omega_d$) and the feedback gain ($g$) are swept (see appendix A for details regarding the experimental setups). The experiments reported here, are performed for $\theta = 0$.
\subsection{\label{sec:level2}Small amplitude libration}
\indent\indent\ A first demonstration of a librator is performed for zero detuning , i.e. $\delta = 0$, and a drive voltage of 400 mV placing the device well within the nonlinear regime as shown in Fig.~2(a). The libration amplitude $R_L$ is observed as the gain of the feedback loop is increased, shown in Fig.~2(b). When the loop gain crosses a critical threshold, $\gamma_L$ becomes negative and the system exhibits a libration limit cycle generated via a Hopf bifurcation. The system is now a librator, Figs.~2(c) to 2(e). Despite the sharp emergence of the limit cycles, as their amplitude increases beyond the onset threshold the scaling more closely resembles the well known square root relation, the libration amplitude versus gain data can be fitted to give a scaled nonlinear damping term of $\beta = 2.62$, also shown along with numerical simulations in Fig.~2(b) \footnote{The shape and scale of the numerically obtained points (red dots) in the inset of Fig.~2(b) is accurate, however, the onset of self oscillation obtained from simulation was centered around a gain value of $g = 32$, this is due to the coarse calibration of gain and nonlinear damping from the experimental data. For visual clarity, the numerical points were shifted to overlap with the experimental data.}.\\
\indent\indent\ Furthermore, Figs.~2(d) and 2(e) show an interesting transition as the limit cycle grows to encompass the origin of the rotating frame phase space. The rotations around the origin of the phase space plane determine the phase in the laboratory frame, if the librator limit cycle does not encompass the origin then the average phase of the system as seen in the laboratory frame is unchanged, i.e. $\langle \phi \rangle = \phi_0$, Fig.~2(d). However, when the limit cycle does encompass the origin the phase of the system starts to rotate, i.e. $\langle \phi \rangle \neq \phi_0$, this free running phase changes the mean frequency of the system, i.e. the mean frequency is no longer that of the drive ($\omega_d$), as shown in Fig.~2(e). This effect is further confirmed by the libration sidebands amplitude overtaking the driven amplitude as shown in the spectral responses in Figs.~2(c) to 2(e). Strictly speaking this transition is not a bifurcation and has been already identified, although not labeled, in the context of strongly forced oscillators \cite{holmes1978bifurcations,levina1986analysis,pikovsky2003synchronization,pikovsky2000phase}, and more recently in the Kuramoto model \cite{wright2020missed}. In other contexts the case of zero average phase is referred to as libration while the case of a free running phase is referred to as rotation \cite{strogatz2018nonlinear,alexeeva2000impurity}. Note that this nomenclature is not prevalent \cite{lenz2007classical,nakajima1994orientational,rost1992pendular,friedrich1995alignment,lima1991fast}, but will be used here to distinguish the two regimes. Furthermore, throughout the text we identify this librator-to-rotator transition by the  acronym ``totoro".\\
\indent\indent\ Subsequently, we sweep the drive frequency, $\omega_d$, while maintaining a constant driving force of $F_d = 400$ mV, and determine the librator's frequency around the onset of the Hopf bifurcation for each of the drive frequencies. This collection of $\omega_L$ is plotted as a function of the detuning parameter, $\delta$, as shown in Fig.~3 along with values calculated from Eq.~(4).\\
\begin{figure*}[th]
	\graphicspath{{Figures/}}
	\includegraphics[width=170mm]{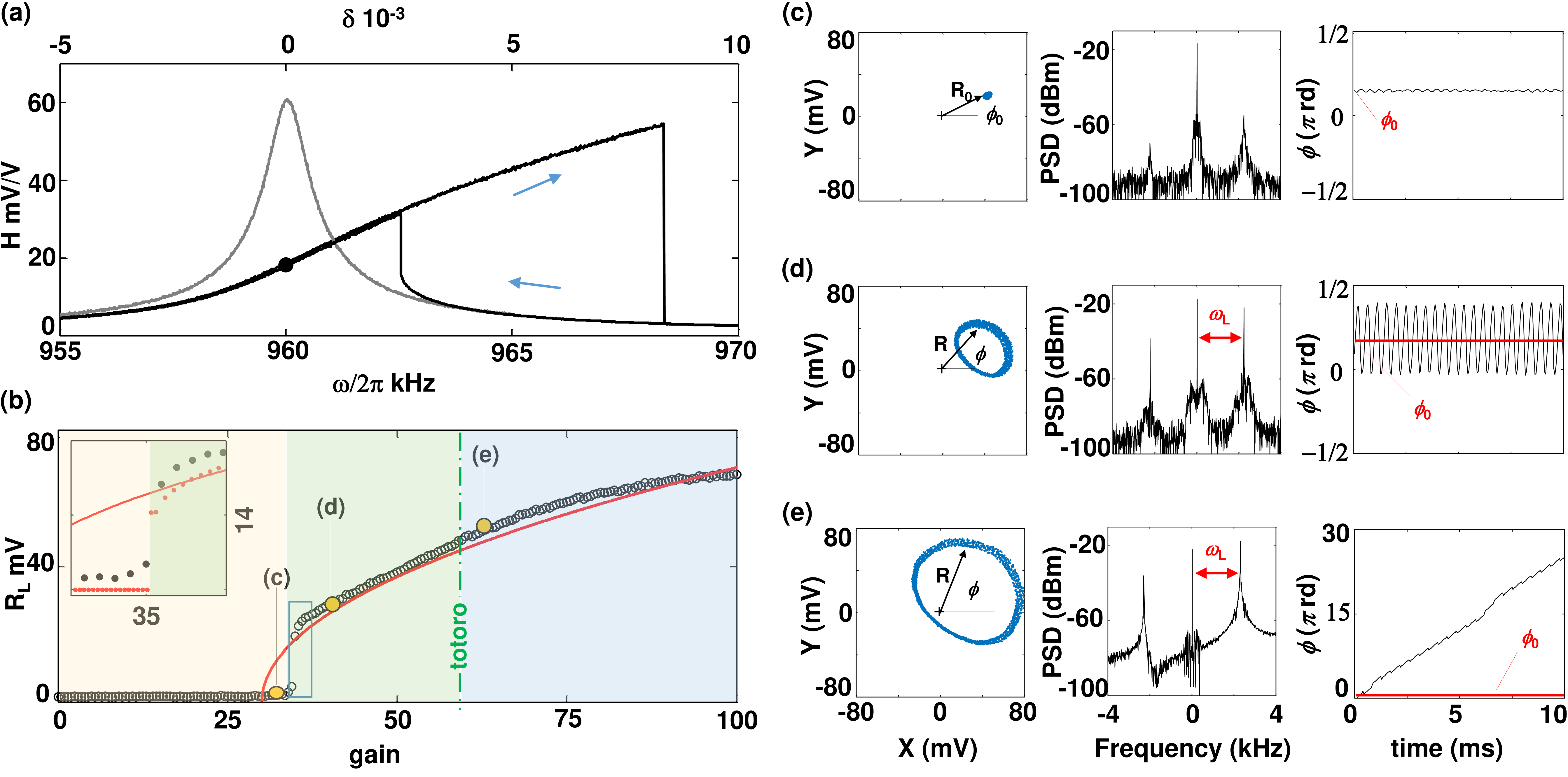}
	\caption{(a) Measured spectral response of the MEMS device for the linear ($F_d = 20$ mV), and the nonlinear bistable ($F_d = 400$ mV) regimes, in grey and black respectively, and $\theta=0$ for both. \emph{H} denotes the relative amplitude response, expressed in mV measured per V drive. The blue arrows show the sweep direction for the lower and upper branches, and the vertical line at $\delta=0$ indicates the parameters around which the librator is created. (b)  Onset and amplitude of libration limit cycles as the feedback loop gain is increased, different background colors indicate the operating regime. Before the Hopf bifurcation (yellow), after the Hopf bifurcation and before the totoro transition (green), and post totoro transition (blue). Experimental data points are shown as black circles, and a square root dependence is plotted to show the onset of the supercritical Hopf bifurcation around a gain of 33 (solid red line). 
	The totoro transition is highlighted. The inset shows a zoom in of the blue rectangle, where the experimental data (black dots) and Eq.~(2)-based simulations (red dots) both show the onset of the limit cycle. The red trace in the inset is the same square root relation shown in the main plot. Details of subthreshold (c), librator (d), and rotator (e) operation. Panels show the rotating frame trajectories (left panels) as the gain is increased, first a limit cycle is created around the initial fixed point (d), the system is a librator. Then the orbit encompasses the phase space origin (e), the system is a rotator. The spectral response (mid panels) demonstrates the asymmetric nature of the librator, also clearly visible is the sideband overtaking the center frequency component as the system crosses the totoro transition. The libration frequency $\omega_L$ is indicated in red in the spectral responses. The phase  (right panels), as extracted from the phase space trajectories, shows the steady state phase $\phi_0$ for subthreshold operation (c), the oscillating phase component with an average phase of $\phi_0$ for the librator (d), and the unbounded phase for the rotator case (e). In all three cases the steady state average phase is indicated (red).}
\end{figure*}
\begin{figure}
	\graphicspath{{Figures/}}
	\includegraphics[width=85mm]{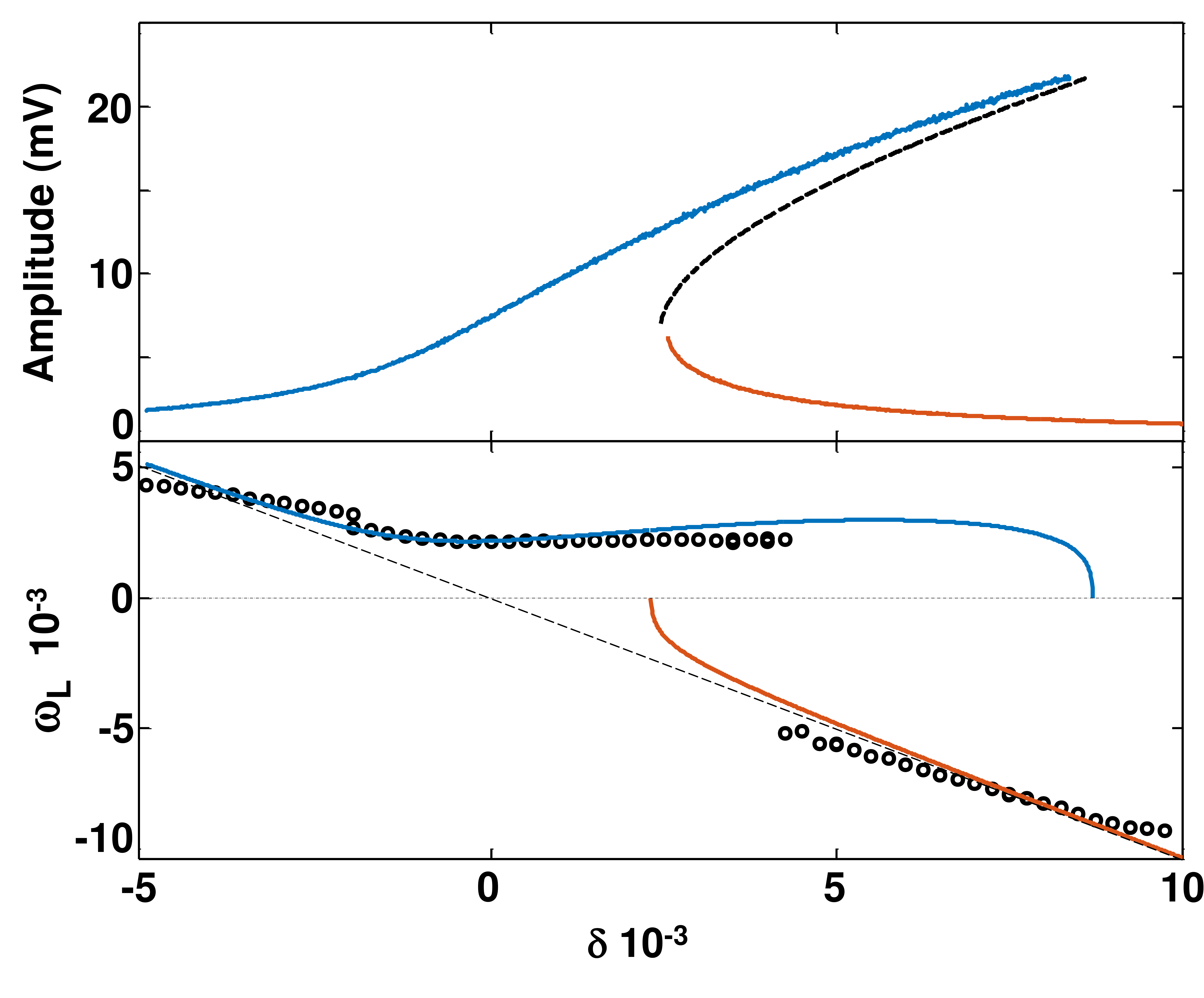}
	\caption{Dependence of the libration frequency on detuning. The nonlinear Duffing response (top panel) is shown as a visual reference, we designate the upper and lower measured branches by the blue and red traces respectively, and the unstable branch, calculated from the fitted device parameters, by the dashed black line. The measured libration frequency at the onset of the Hopf bifurcation (bottom panels). The measured values are shown in black circles, and the solid lines correspond to the analytically calculated libration frequency as given by Eq.~(4), equally shown in blue and red for the upper and lower branches, respectively. Note that for large detunings the libration frequency practically follows the linear relation $\omega_{Libration} = -\delta$, shown as the black dashed line. The areas with no experimental data corresponds to parameters where the system was not stable enough to perform the measurements.}
\end{figure}
\indent\indent\ Particularly interesting is the bistable interval. Since two possible steady state solutions exist, then there equally exist the possibility to induce two libration limit cycles around each one of those solutions, although not simultaneously. As can be seen in Fig.~3, the libration frequency drops to zero around the saddle-node bifurcations, where one of the stable fixed points and the saddle point collide. Whether within or outside the bistable region, the small amplitude librator frequency agrees relatively well with the analytical calculations further confirming the linearization in Eq.~(4).
\subsection{\label{sec:level2}Large amplitude behaviour
}
\indent\indent\ The possibility to generate two limit cycles centered around each of the steady state solution branches merits an in depth look at the large amplitude response of the librator. For one, libration limit cycles around the low-amplitude branch (LB) and the high-amplitude branch (HB) orbit their respective fixed points in opposite directions, clockwise and counter-clockwise, respectively, as can be seen from the experimental data in Fig.~4(a). Indicating that the LB librator has a dominant negative frequency component within the rotating frame, while the HB librator has a dominant positive frequency component within the rotating frame. Indeed, previous work found that driven or even stochastic librations tend to show a strong asymmetry depending on detuning \cite{houri2020generic,huber2020spectral,ochs2020multiphoton}.\\
\indent\indent\ At larger amplitudes, both the HB and LB limit cycles are bounded by homoclinic bifurcations, i.e. limit cycles with infinitely long periods that pass through the saddle point. As the gain of the feedback loop is increased, the limit cycles approach the homoclinics and as a consequence their frequency reduces as their amplitude increases. Thus librators exhibit a very strong nonlinearity whereby their frequency starts with $\omega_L$ as given by Eq.~(4) for $A_L\approx 0$, and ends with $\omega_L = 0$ for $A_L= A_{Homoclinic}$. Experimental and numerical demonstration of this slowing down near the homoclinics is shown in Fig.~4(b), where the libration (or rotation) frequency is plotted as a function of the maximum distance between a limit cycle trajectory and the repsective fixed points (max($R_L$)).\\
\indent\indent\ As the gain is increased beyond the homoclinic bifrucations, the limit cycles, whether originally orbiting the HB or LB, transition to a new regime, one whose orbit now encompasses all three fixed points and rotates in a counter-clockwise fashion, equally shown in Fig.~4(a). This additional limit cycle is made possible by the fact that the three fixed points have a cumulative index number of 1, and would not have been possible in a two-dimensional phase space exhibiting only two fixed points \cite{takens2001forced}. These wide limit cycles are in fact always rotators, whereas prior to the homoclinic bifurcation the limit cycles can be either librators or rotators. Furthermore, they demonstrate the same scaling behaviour as their progenitor limit cycles, i.e. slowing down as they approach the homoclinic, as shown in Fig.~4(b).\\
\indent\indent\ The transition between the different limit cycle regimes underlines limitations in the librator model. For one, it is important to keep in mind that the librator dynamics, whether small or large amplitude, are defined around a DC component. Therefore, if the system is in some condition which destabilize the steady state, i.e. a transition from the high branch to the low branch, then the transient is not accounted for by the current model. Furthermore, the librator, as implemented by the feedback loop shown in Fig.~1(b), revolves around a DC component which we have approximated by the steady state solution(s) to the driven Duffing equation. When the librator transitions to the large rotation orbits, such approximation is no longer valid as the difference between the DC component (the mean value of $A_L$ over an orbit) and the steady state solution is significant. Indeed, for the large rotation orbits we re-write the dynamics equation into one that is independent of the steady state solutions, which reads (see Appendix B for detailed derivation)\\
\begin{multline}
i\dot{A}_L -\delta A_L +\frac{i\gamma}{2}A_L +\frac{3\alpha}{8}\vert A_L\vert^2A_L + \frac{i\beta}{8}\vert A_L\vert^2A_L \\
= \frac{1}{2}(F_d + \frac{ig}{4}(A_Le^{-i\theta} -A_L^*e^{i\theta}))
\end{multline}\\
\indent\indent\ The piecewise model is a consequence of this limitation, where Eq.~(2) is used for libration around the fixed points (with the respective parameters accounted for), and Eq.~(5) is used for the large amplitude orbits.\\
\indent\indent\ On a side note, the combined presence of a harmonic drive and a limit cycle may be confused with the case of a forced oscillator, however the two represent largely distinct dynamics and bifurcation diagrams. The fundamental difference between the two being that in the case of the librator the driving force creates a certain phase space topology which is largely unchanged by the limit cycle, whereas in the case of a forced oscillator, the limit cycle and the driving force interact to create the topology. As a consequence the two systems exhibit widely differing behaviour. For one a forced oscillator locks its frequency and phase in response to weak external forcing, whereas the librator, virtually by definition, does not. If the external forcing is highly detuned, a forced oscillator can experience de-synchronization leading to phase slips via a SNIC bifurcation (Saddle Node on an Invariant Circle), whereas the librator only changes its frequency as the detuning is changed. Furthermore, under the effect of strong external forcing, non-isochronous oscillators exhibit a highly complicated bifurcation diagrams \cite{holmes1978bifurcations,levina1986analysis,pikovsky2003synchronization,mayol2002class} with the potential to generate chaos \cite{kawaguchi1984new,lee1993bistability,simpson1994period}, whereas the librator, as stated, does not produce a steady state chaotic output. Incidentally, one feature that is in common to both forced oscillators and librators is the possibility to observe totoro transitions, since it can be argued that such transitions are common to multi-frequency dynamical systems \cite{pikovsky2000phase,wright2020missed}.\\
\begin{figure}
	\graphicspath{{Figures/}}
	\includegraphics[width=85mm]{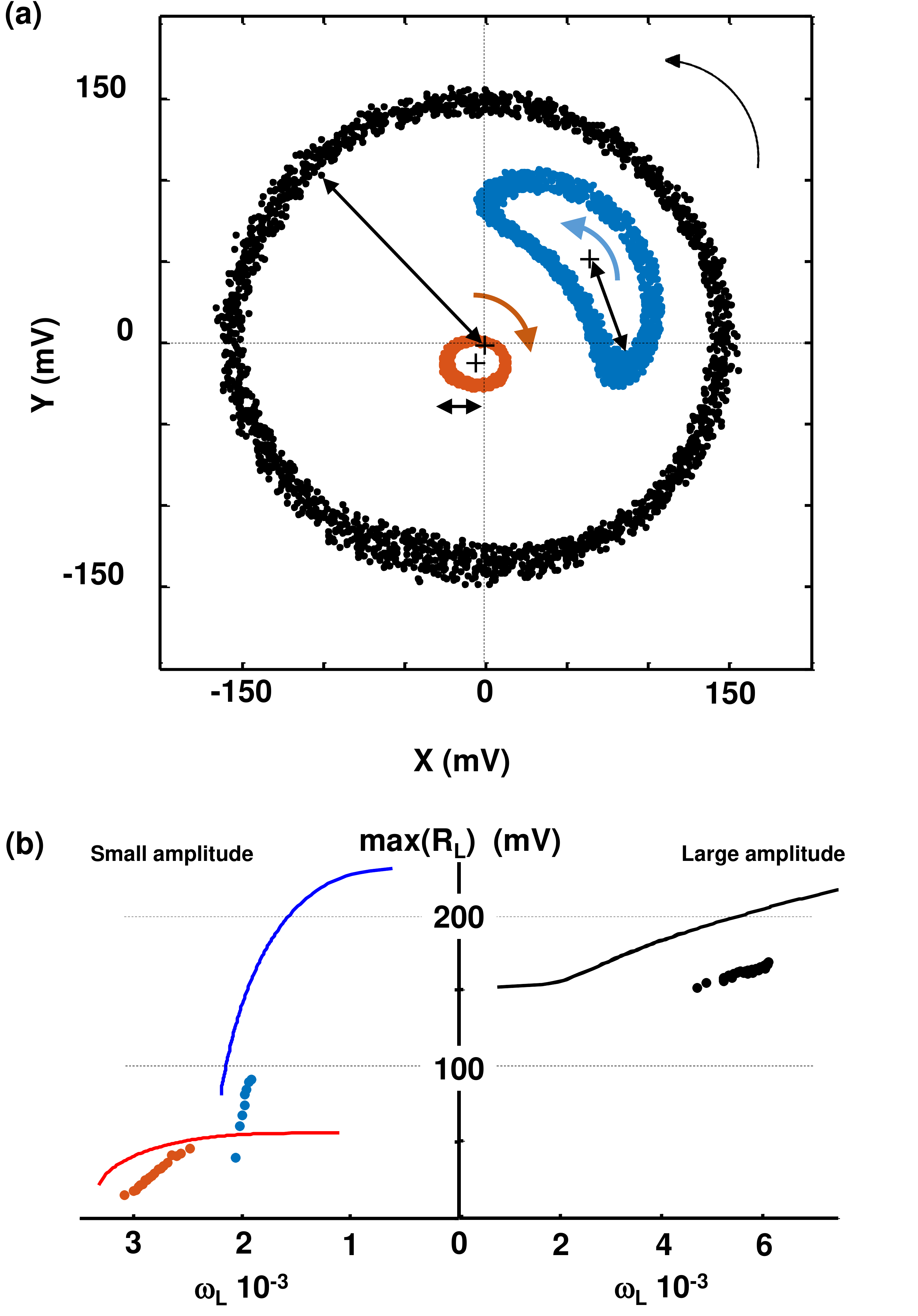}
	\caption{(a) Libration and rotation orbits obtained for $F_d=0.6V$. Orbiting counter-clockwise for the high amplitude branch (blue), and the large rotator regime (black), and clockwise around the low amplitude branch (red). The black double-sided arrows indicate the maximum libration distance with respect to the fixed points (in the case of the large amplitude rotation it is with respect to the origin). The black and blues traces are obtained for $\delta = 2.6\times 10^{-3}$ (but for different amplifier settings), while the red trajectory is obtained for $\delta = 3.6\times 10^{-3}$. (b) Scaling of libration or rotation frequency as a function of the distance between the homoclinic and the maximum libration distance ($A_{Homoclinic}-$max($R_L$)). The left panel represents the libration or rotation frequency prior to the crossing of the homoclinic bifurcation, with the high-branch and low-branch data shown in blue and red, respectively (circles for experimental data, and solid lines for simulation). The right panel shows the scaling post the homoclinic bifurcation. The mismatch between simulation and measurements indicates experimental frequency drift and amplitude calibration drift.}
\end{figure}
\section{\label{sec:level1}Conclusions}
\indent\indent\ Several exciting prospects for further investigation of librators are possible. For one, the use of nonlinear resonators with higher order nonlinearity, say quintic nonlinearity \cite{kacem2015overcoming,samanta2018tuning,huang2019frequency}, implies the possibility of even more distinct limit cycles and homoclinic bifurcations within the same phase space, which is an outlook of practical and fundamental interest \cite{arnol1977loss,kuznetsov2013visualization,leuch2016parametric}. Indeed, even in the system presented in this work a rigorous account of the existence and number of limit cycles (both stable and unstable ones) was not fully given, and these questions remain to be addressed on a theoretical and numerical level.\\
\indent\indent Furthermore, that a librator maybe synchronized by the application of a weak external forcing, in a manner similar to the way oscillators can be synchronized, is worth investigating. And the formation of a librator network, potentially within a single multi-mode device \cite{houri2020demonstration} could be of great practical importance.\\
\indent\indent\ On an experimental note, it may be possible to produce a feedback loop-free librator, in which thermomechanical back-action can theoretically trigger a Hopf bifurcation in a driven high quality factor nanomechanical resonator \cite{dykman2019resonantly}, although this remains to be proven experimentally. 
On another hand, if one accepts that $F_d \gg f$ (below the totoro transition) then the lock-in amplifier can be replaced with an envelope detector, thus greatly simplifying the experimental setup.\\
\indent\indent\ In summary, this work introduced the ``librator" as a new dynamical mode of operating nonlinear MEMS devices, in which a quasi-periodic output is generated through the creation of limit cycles within the rotating frame of a driven nonlinear MEMS resonator. These limit cycles do not change the global topology of the rotating-frame which is created by the driving force, but are rather constrained by it. Different types of limit cycles are observed, along with homoclinic bifurcations. These bifurcations do not induce chaos as the system is contained within a two-dimensional phase space. 
Interestingly, the dynamics presented here can be applied to other physical implementation of weakly nonlinear and weakly damped resonators, such as optical and superconducting ones.\\
\begin{acknowledgments}
 \indent\indent\ The authors would like to thank prof. L. Minati for useful discussions.
\end{acknowledgments}

\appendix

\section{Experimental setups and procedures}
\indent\indent\ The Duffing parameter is characterized using a series of nonlinear frequency response curves, which are obtained using a Zurich-Instrument lock-in amplifier (HF2LI) under a -0.5 $V_{DC}$ bias. The negative bias is applied to avoid electrical nonlinearities in the metal-semiconductor contact \cite{yamaguchi2017gaas}. The fits, some of which are shown in Fig.~6, follow the procedure detailed in \cite{houri2019limit,davidovikj2017nonlinear}.\\
\indent\indent\ Throughout the measurement period, a very slow (day time scale) frequency drift is observed in the device, this is accounted for by performing a spectral response sweep before each measurement run. This slow frequency drift only affects $\omega_0$ and $\delta$, but has no impact on the quality factor and the nonlinear terms.\\
\begin{figure}
	\graphicspath{{Figures/}}
	\includegraphics[width=85mm]{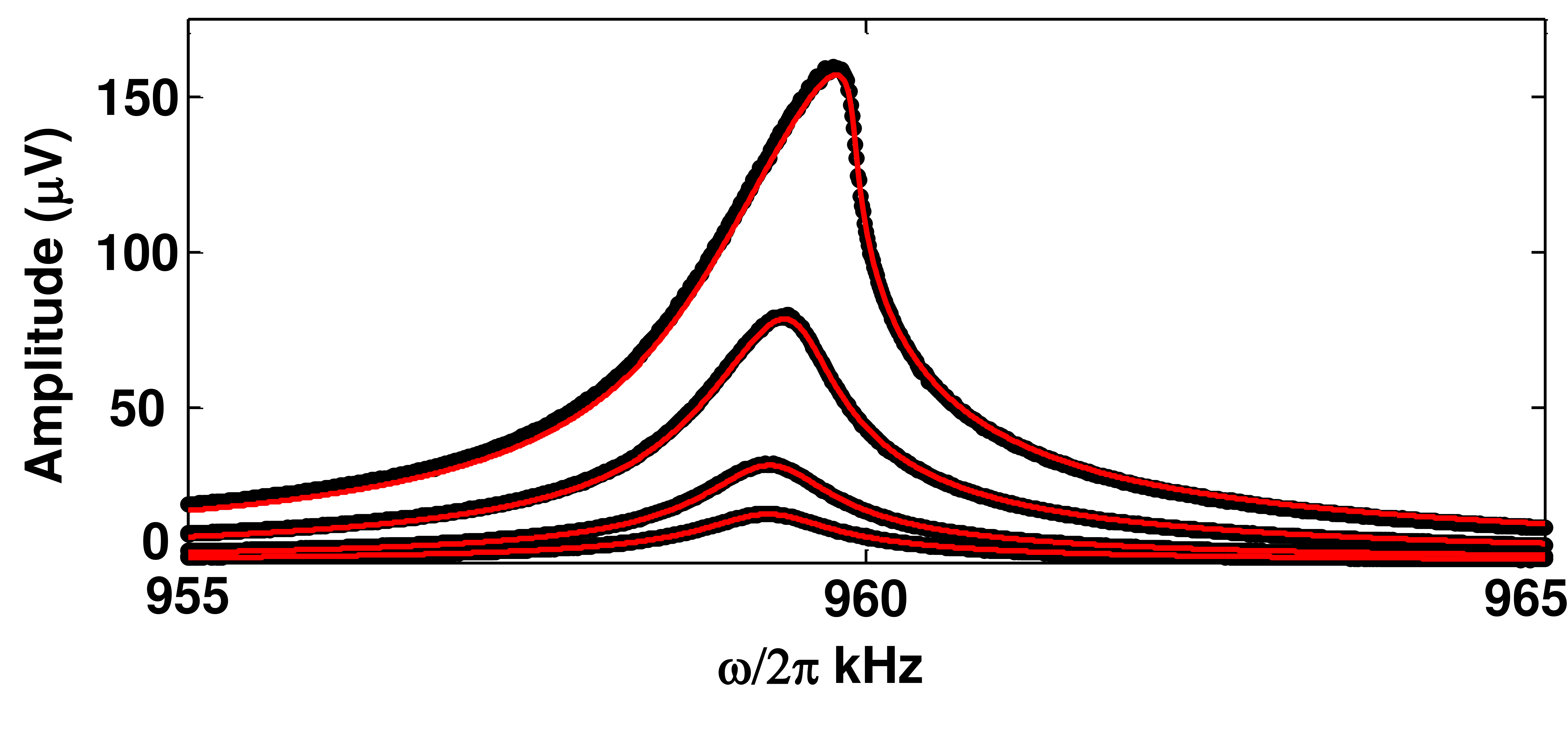}
	\caption{Fits of the nonlinear resonance response. A few examples of the experimental nonlinear responses (black circles) and their fits (red lines) for drive amplitudes of 20, 50, 100, and 150 mV for the curves with increasing amplitudes, respectively.}
\end{figure}
\indent\indent\ We implement the feedback loop as shown in Fig.~7 using the following instruments, a Neaoark LDV (Neoark Corporation) with a 100 MHz bandwidth and a 10 m/s/V sensitivity is used to measure the device.We use a lock-in amplifier (SR844, SRS) to down-convert the output from the LDV, and subsequently a filter-amplifier (NF37627, NF corporation) to filter and amplify the X-quadrature from the output of the lock-in. A vector signal analyzer (VSA, HP89410A, Keysight) equally samples the output of the LDV. We use the output from the filter-amplifier to perform a Double-Sideband Transmitted Carrier Amplitude Modulation (DSB-TC AM) using a waveform generator (WF1974, NF corporation). Because the modulation depth is limited to 100\% this imposes a limit on the feedback amplitude such that $\vert\emph{f}\vert = F_d$.\\
\indent\indent\ 
To obtain Fig.~4 a Double-Sideband Suppressed Carrier (DSB-SC) was used, with an additional drive tone generated using an independent channel on the signal generator, this was done to overcome the modulation depth limitation of the DSB-TC configuration. Furthermore, the nature of the used electronic filter is such that a high feedback loop gain can potentially ring the filter itself, i.e. turn the filter into an oscillator. The output from the system is closely monitored during measurements, to ensure that such behavior does not occur.\\
\begin{figure}
	\graphicspath{{Figures/}}
	\includegraphics[width=85mm]{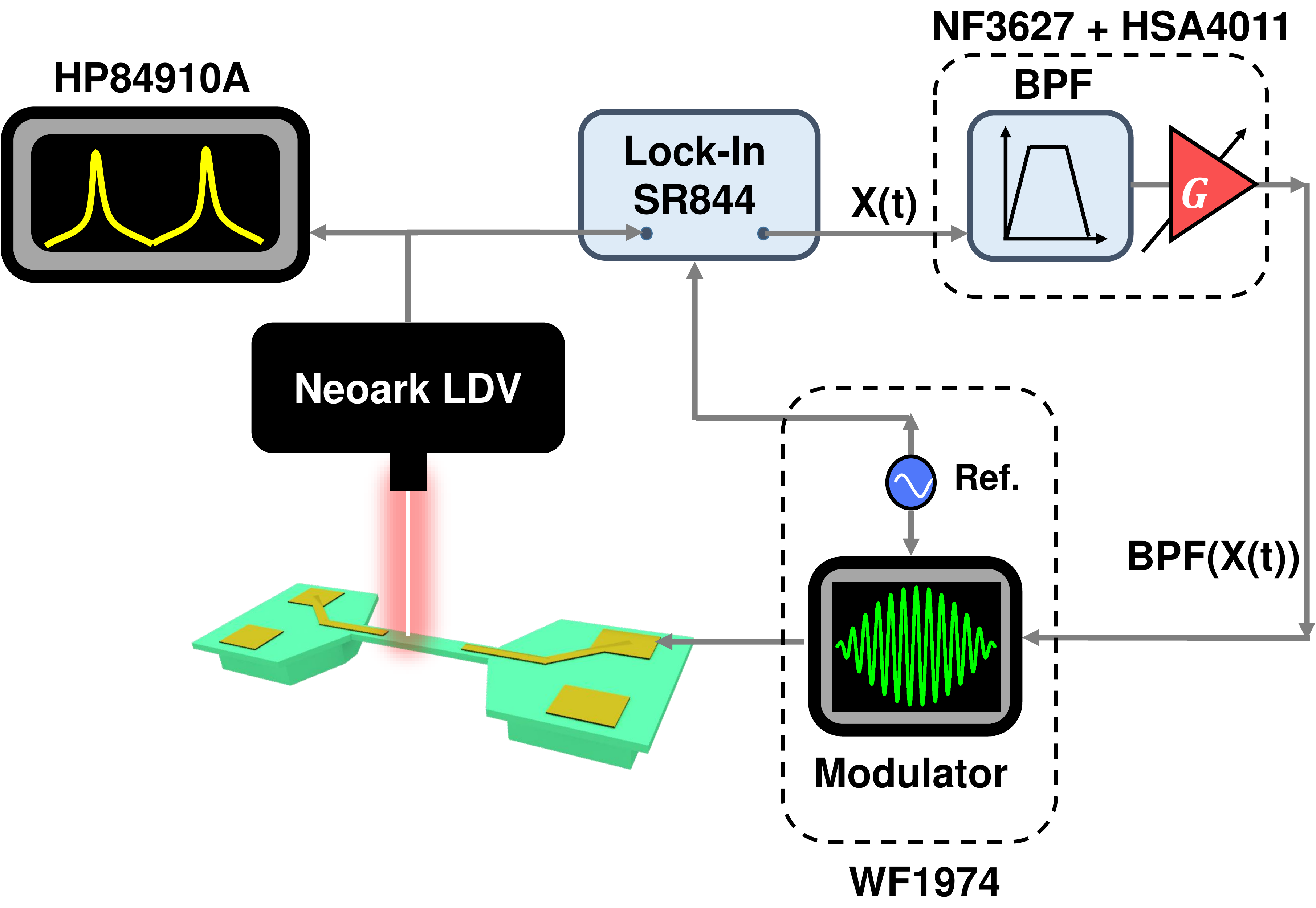}
	\caption{Schematic representation of the implementation of the librator feedback loop.}
\end{figure}
\section{Derivation of librator dynamics}
\indent\indent\ We start by deriving a closed form for the term $f$ in Eq.~(1). Using the rotating frame approximation (RFA), i.e. $x(t) = \frac{1}{2}(Ae^{i\omega t} + A^*e^{-i\omega t})$, the output of the laser Doppler vibrometer (LDV) is then given as\\
\begin{equation}
\dot{x}(t) = \frac{i\omega}{2}(Ae^{i\omega t}-A^*e^{-i\omega t}) + \frac{1}{2}(\dot{A}e^{i\omega t} + \dot{A}^*e^{-i\omega t})
\end{equation}\\
\indent\indent\ After multiplying the LDV output with the reference tone, i.e. $\cos(\omega t)$, we obtain\\
\begin{multline}
S(t) = \dot{x}(t)\cos(\omega t + \theta) = \\
\frac{i\omega}{4}(Ae^{i(2\omega t+\theta)} + Ae^{-i\theta} -A^*e^{i\theta} - A^*e^{-i(2\omega t + \theta)}) +\\ \frac{1}{4}(\dot{A}e^{i(2\omega t+\theta)} + \dot{A}e^{-i\theta} + \dot{A}^*e^{i\theta} + \dot{A}^*e^{-i(2\omega t + \theta)})
\end{multline}\\
where $\theta$ is an arbitrary phase difference.\\
\indent\indent\ After removing the high frequency components, as a result of the band-pass filter (BPF), Eq.~(B2) reduces to\\
\begin{multline}
S(t) = \frac{i\omega}{4}(Ae^{-i\theta} -A^*e^{i\theta}) +\frac{1}{4}(\dot{A}e^{-i\theta} + \dot{A}^*e^{i\theta})
\end{multline}\\
\indent\indent\ We consider that $\omega A\gg\dot{A}$, and thus drop the second term from the above equation, thus giving\\
\begin{equation}
S(t)  \approx \frac{i\omega}{4}(Ae^{-i\theta} -A^*e^{i\theta})
\end{equation}\\
\indent\indent\ Since we decomposed the complex amplitude into a steady state DC component, and a libration AC component, i.e. ${A}(t) = {A_{DC}} + {A_{AC}}(t) = {A_{0}} + {A_{L}}(t)$, the BPF also removes the DC component from the signal thus reducing $S(t)$ to\\
\begin{equation}
\omega S_{ac}(t) = \frac{i\omega}{4}(A_Le^{-i\theta} -A_L^*e^{i\theta})
\end{equation}\\
\indent\indent\ The signal is then amplified by a factor $g$ and possibly phase-shifted to give $f=gS_{ac}$, we consider that any additional phase shift can still be contained within the $\theta$ term, to give\\
\begin{equation}
f = g\omega S_{ac}(t) = \frac{ig\omega}{4}(A_Le^{-i\theta} -A_L^*e^{i\theta})
\end{equation}\\
\indent\indent\ We rewrite Eq.~(1) as\\
\begin{multline}
{\ddot{x} + {(\gamma + \beta x^2)}\dot{x} + \omega_{0}^2x + \alpha x^3}\\
= {{(F_d + g\omega S_{ac}(t))}\cos(\omega_dt)}\\
\end{multline}\\
\indent\indent\ We apply the standard RFA approximations to Eq.~(B7), where we introduce $x(t) = \frac{1}{2}(Ae^{i\omega t} + A^*e^{-i\omega t})$ into Eq.~(B7) and keep only the $\omega_d t$ terms of first order to give\\
\begin{equation}
\begin{split}
\ddot{x}~\approx&~\frac{-\omega_d^2}{2}Ae^{i\omega_d t}+\frac{i\omega}{2} \dot{A} e^{i\omega_d t}\textrm{,}\\
\gamma\dot{x}~\approx&~\frac{i\omega_d\gamma}{2}{A} e^{i\omega_d t}\textrm{,}\\
\alpha x^3~\approx&~\frac{3\alpha}{8}{AA^*A} e^{i\omega_d t} \textrm{,}\\
\beta x^2\dot{x}~\approx&~\frac{i\beta}{8}{AA^*A} e^{i\omega_d t} \textrm{,}\\
 \omega_0^2x~\approx&~\frac{\omega_0^2}{2}Ae^{i\omega_d t} \textrm{,}\\
 \omega_d^2~\approx&~\omega_0^2(1+2\delta) \textrm{.}\\
\end{split}
\end{equation}\\
\\
\indent\indent\ Thus Eq.~(B7) becomes (in non-dimensional form)\\
\begin{multline}
i\dot{A} -\delta A +\frac{i\gamma}{2}A +\frac{3\alpha}{8}AA^*A + \frac{i\beta}{8}AA^*A \\
= \frac{1}{2}(F_d+gS_{ac}(t))
\end{multline}\\
\\
\indent\indent\ If the libration motion is centered around one of the two branches, then we approximate the DC component to the steady state solution of a forced Duffing, we split the complex amplitude into two equations respectively\\
\begin{equation}
(-\delta+\frac{3\alpha}{8}\vert A_0\vert^2)A_0 + i(\frac{\gamma}{2} + \frac{\beta}{8}\vert A_0\vert^2)A_0 = \frac{F_d}{2}
\end{equation}\\
and,\\
\begin{multline}
i\dot{A}_L -\delta A_L +\frac{i\gamma}{2}A_L\\
+\frac{3\alpha}{8}(2\vert A_0\vert^2A_L + A_0^2A^*_L + 2\vert A_L\vert^2 A_0 + A_0^*A_L^2 + \vert A_L \vert^2A_L)\\
+ \frac{i\beta}{8}(2\vert A_0\vert^2A_L + A_0^2A^*_L + 2\vert A_L\vert^2 A_0 + A_0^*A_L^2 + \vert A_L \vert^2A_L)\\
= \frac{ig}{8}(A_Le^{-i\theta} - A_L^*e^{i\theta})\\
\end{multline}\\
\indent\indent\ Equation~(B10) is the standard RFA of a driven Duffing, while by rearranging Eq.~(B11) we obtain Eq.~(2) from the main text.\\
\indent\indent\ If on the other hand the libration is in the regime of the large amplitude limit-cycle where the orbit encircles all three fixed points, then due to the almost circular shape of the orbit and the fact that it is nearly centered around the origin of the phase-space we approximate the DC component to zero, i.e. $A_{DC} \approx 0$, thus leading for Eq.~(B9) to be rewritten as\\
\begin{multline}
i\dot{A}_L -\delta A_L +\frac{i\gamma}{2}A_L +\frac{3\alpha}{8}\vert A_L\vert^2A_L + \frac{i\beta}{8}\vert A_L\vert^2A_L \\
= \frac{1}{2}(F_d + \frac{ig}{4}(A_Le^{-i\theta} -A_L^*e^{i\theta}))
\end{multline}\\
\bibliography{apssamp}

\end{document}